\documentclass{mem}
\usepackage{natbib}\usepackage{txfonts}\usepackage{balance}
\usepackage{graphicx}
\usepackage[a4paper]{hyperref}
\idline{000}{000}
\begin{document}
\def\teff{$T\rm_{eff }$}
\def\kms{$\mathrm {km s}^{-1}$}

\title{
Close frequencies in pulsating stars: common and mysterious!
}

   \subtitle{}

\author{
M. \,Breger\inst{1}
\and A. A. \, Pamyatnykh\inst{1,2,3}
          }

  \offprints{M. Breger}

\institute{
Astronomisches Institut der Universit\"at Wien, T\"urkenschanzstr. 17, A--1180 Wien, Austria
\email{michel.breger@univie.ac.at}
\and
Copernicus Astronomical Center, Bartycka 18, 00-716 Warsaw, Poland
\and
Institute of Astronomy, Pyatnitskaya Str. 48, 109017 Moscow, Russia
}

\authorrunning{Breger and Pamyatnykh}

\titlerunning{Close frequencies in pulsators}

\abstract{
Amplitude and phase variability are commonly found in many different types of pulsating
stars. This suggests a common, presently unknown physical origin. We have examined the phenomenon in several $\delta$~Scuti stars with extensive data and find the beating of close frequencies to be responsible. This is demonstrated for the star FG Vir by testing the
relationship between the observed amplitude and phase variations. Most close frequency pairs are situated near the observed or theoretically predicted frequencies of radial modes. The large number of detected close frequencies excludes the possibility of accidental frequency agreements.
\keywords{Stars: $ \delta$ Sct -- Stars: variables: general -- Stars: oscillations - Stars: individual: FG Vir}
}
\maketitle{}

\section{Introduction}

The search for the astrophysical origin of the observed amplitude and phase variability associated with stellar pulsation concerns many different stellar pulsators in vastly different stages of stellar evolution. We refer to this phenomenon as the Blazhko Effect, which was noticed by Blazhko (1907) in the RR Lyrae star RW Dra. These variations are quite common in RR Lyrae stars, e.g., Szeidl (1988) mentions an
occurrence of 20 - 30 \% in the RRab stars. However, a large number of different types of pulsators in different
stages of evolution also show amplitude and phase variability on non-evolutionary time scales. Examples
are classical cepheids (e.g., Breger 1981), sdB stars (e.g., Kilkenny et al. 1999), $\beta$  Cephei stars
(Lehmann et al. 2001), White Dwarfs (e.g., Handler et al. 2003), and $\delta$ Scuti stars.

Regrettably, the physical origin of these variations are not yet known. This might be caused
by the lack of extensive observational data to distinguish between different hypotheses.

\section{Origin of amplitude variability in $\delta$~Scuti stars}

\begin{figure*}
\centering
\includegraphics*[bb=68 294 768 755,width=135mm,clip]{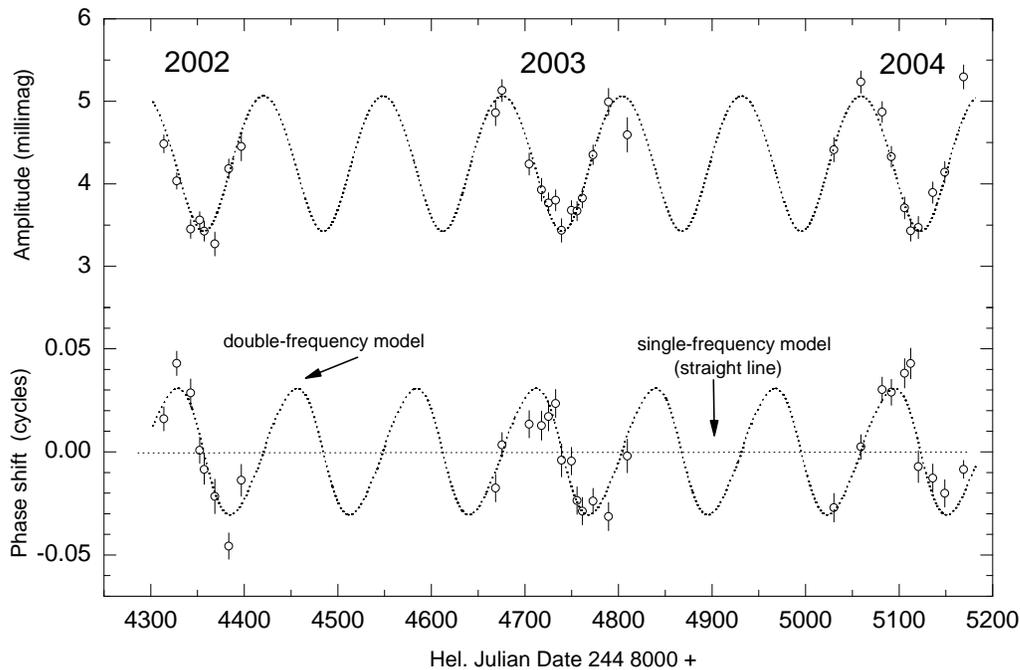}
\caption{\footnotesize
Amplitude and phase variations of the 12.154 c/d frequency in FG Vir. The open circles
represent the observations, while the dotted line is the two-frequency fit. The diagram shows that
the variations can be explained by two independent frequencies separated
by 0.0078 c/d and an amplitude ratio of 0.19. Note the correct signature
of beating: the time of minimum amplitude corresponds to the time of average phasing
and the most rapid phase change. Also note the excellent agreement between the
different years.}
\end{figure*}

%
%

$\delta$ Scuti variables are excellent candidates to examine the phenomenon because of the available extensive studies of selected stars. These multiperiodic stars are in the main-sequence or immediate post-main-sequence stage of evolution. Similar to the situation in White Dwarfs, the amplitude variability can be very extreme: a star may change its pulsation spectrum to such an extent as to appear as a different star at different times, at least in its pulsational behavior. However, detailed analyses of the $\delta$ Scuti star 4 CVn showed that the modes do not completely disappear, but are still present at small amplitudes (Breger 2000).

For the majority of the well-studied $\delta$ Scuti stars, frequency pairs with frequency
separations less than 0.1 c/d can be seen in the power spectra. These pairs may be
a reflection of close frequencies, amplitude variability of a single mode or observational errors.
It is possible to distinguish between true amplitude variability and beating of two close frequencies:
In the Fourier spectrum, two close frequencies beating with each other appear as a single frequency with variable amplitude and phase. The resulting amplitude
and phase variations are mathematically related. The most extreme and easily recognizable
situation occurs when two close frequencies have the same amplitude: this leads to a half a cycle phase shift at the time of minimum amplitude of the visible (single) frequency. In general, the largest phase change (of an assumed single frequency) occurs at the time of minimum amplitude. Therefore, it is possible to separate beating from true amplitude variability by studying the relationship between the amplitude and phase variations of an assumed single frequency. For the star BI CMi, Breger \& Bischof (2002) showed from detailed phase-amplitude tests that
the amplitude variability was the result of beating of separate modes with close frequencies.

We can now apply the test of the close-frequency hypothesis to a number of modes found in the extensive new photometric data of FG Vir (Breger et al. 2005). This data set provides coverage over 98, 160, 165 nights in the years 2002, 2003 and 2004, respectively. Let us now examine the 12.154 c/d mode. The following approach was adopted: the photometric data were subdivided into bins each covering approximately one week. For each bin, the best amplitude and phasing was computed assuming a single frequency. (More details concerning the use of data obtained through different filters as well as the elimination of effects from the other modes present in the star can be found in a paper presently in preparation to be sent to MNRAS.) Fig. 1 shows the variations in amplitude and phase for the three years. In this figure, the uncertainties were calculated by using the standard relations for calculating amplitude and phasing uncertainties (Breger et al. 1999).

The variations are similar in all three years with a beat period of 128d.
Another important result is that the phase changes are coupled to the amplitude
changes. In particular, minimum amplitude occurs at the time of 'average' phase and
the time of most rapid phase change. As already mentioned earlier, this is an important
signature of beating between two close frequencies. The visual result is confirmed by a two-frequency model with the optimum parameters of frequency, amplitude and phase determined by PERIOD04
(Lenz \& Breger 2005). An excellent agreement is obtained for both the amplitude
and phase changes with the two-frequency model. If we extend the analysis to include the 1995 data, we also find an excellent agreement between the predicted and observed amplitudes and phases of the earlier measurements.

We conclude that the amplitude and phase variations of the 12.154 c/d frequency can be
explained by the beating of two close frequencies. The analysis has been repeated for a number of other pulsation modes in FG Vir. The results are the same: the amplitude and phase variability is caused by beating of close frequencies.

\section{Affinity for radial modes?}

\begin{figure*}
\centering
\includegraphics*[bb=80 374 800 774,width=135mm,clip]{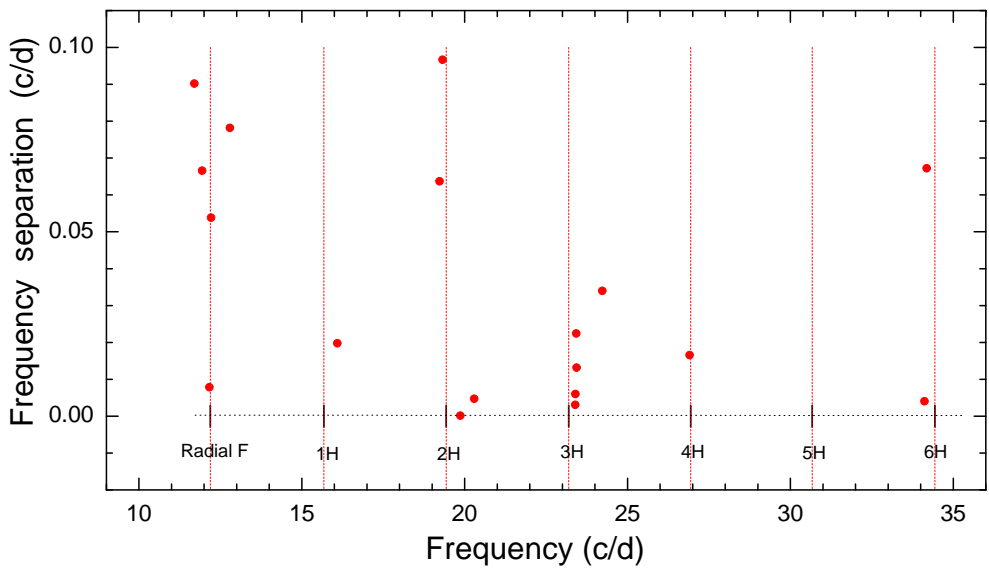}
\caption{\footnotesize
Frequencies of close pulsation modes for all pairs with a separation less than 0.1 c/d.
The y-axis represents the frequency separations of the pairs. The positions of the computed radial
modes are shown. We note that most of the close frequency pairs are situated near the expected frequencies of the radial modes.}
\end{figure*}

Multifrequency analyses of the photometric data of the $\delta$ Scuti star FG Vir coupled with amplitude and phase tests (previous section) show that the star contains 18 sets of close frequencies with
a frequency difference smaller than 0.1 c/d. The frequency distribution of close frequency pairs may contain important information about their physical origin.

In Fig. 2 we have plotted the frequency values at which the close pairs occur. Also shown are the computed frequencies of the radial modes of FG Vir using the model which best fits all the photometric and spectroscopic data and also fits the radial fundamental mode frequency to the observed frequency 12.154 c/d. At the present time, only this mode has a secure detection and identification as radial mode (Breger et al. 2005, Daszynska-Daskiewicz et al. 2005). One more observed mode, 16.071 c/d, was identified by Daszynska-Daskiewicz et al. (2005) also as a radial mode, the frequency is close to the expected value for the first overtone, and an observed close pair occurs here too (16.071 and 16.091 c/d). However, this mode identification has to be confirmed independently with different methods.

The figure shows that the majority of the close frequencies are found near
expected radial modes. This suggests that an accumulation of nonradial modes
around the radial frequencies may occur. The situation is similar to
that observed in RR Lyrae stars (Olech et al. 1999), but important differences
should be stressed. In RR Lyrae stars, radial modes always dominate in the
frequency spectrum. Also, the frequency spectrum of nonradial modes is much sparser and mode trapping is less effective in Delta Scuti stars than in RR Lyrae stars.

 \section{Are the close frequencies accidental?}

Two $\delta$ Scuti stars (BI CMi and FG Vir) have such extensive data coverage
with excellent frequency resolution so that the relationship between the
amplitude and phase variations could be examined in detail. The results
fully support the interpretation of beating between two (or maybe more) close
frequencies. If we consider the fact that more than 75 frequencies have been detected for
FG~Vir with values between 5.7 and 44.3 c/d, the question arises of whether such
agreement could be accidental. In particular, we need to calculate the probability
of accidental agreements.

Let us consider the 12.154 c/d pair with a separation of 0.0078 c/d. If we assume a random
distribution of frequencies and adopt a Poisson distribution, we obtain 0.02 expected pairs.
Of course, the frequencies are not distributed at random. If we adopt a particularly dense region,
viz., the 9.5 to 13. 5 c/d region, we predict 0.03 pairs. For two or more pairs detected in a star,
an explanation of all agreements being caused by accident must be rejected.

We conclude that in the $\delta$ Scuti variables with extensive photometric data, the amplitude and phase variability associated with some pulsation modes are caused by close frequencies, rather than intrinsic amplitude and phase variations. Furthermore, most of the close frequency pairs found in FG~Vir (and in some other stars such as BI~CMi) are not accidental. The question arises whether this result also holds for other types of pulsating variables with known amplitude changes.

\begin{acknowledgements}
This investigation has been supported by the Austrian Fonds zur F\"{o}rderung der wissenschaftlichen Forschung and by the Polish MNiI grant No. 1 P03D 021 28.
\end{acknowledgements}

\end{document}